\documentclass[A4]{article}

\usepackage[]{graphicx}
\usepackage{amsmath, amssymb}

\newcommand{\bq}       {\begin{eqnarray}}
\newcommand{\eq}       {\end{eqnarray}}
\newcommand{\rf}[1]    {(\ref{#1})}

\newcommand{\en}       {\varepsilon}

\newcommand{\lp}       {\left}
\newcommand{\rp}       {\right}
\newcommand{\de}       {\partial}

\newcommand{\nt}       {\noindent}
\newcommand{\fh}       {\frac 1 2}

\begin{document}

\title{\large \bf A model for steady flows of magma-volatile mixtures}

\author{Marco Belan\footnote{Politecnico di Milano, Dipartimento di Ingegneria Aerospaziale, via La Masa 34 - 20156 Milano, Italy, belan@aero.polimi.it}, \hspace{2mm}
Enzo Cataldo\footnote{Universit\'a Milano Bicocca, Dipartimento di Scienze Geologiche e Geotecnologie, Piazza della Scienza 4 - 20126 Milano, Italy}
}

\date{}

\maketitle                   

\centerline{\bf abstract}
{\small
A general one-dimensional model for the steady adiabatic motion of liquid-volatile mixtures in vertical ducts with varying cross-section is presented. 
The liquid contains a dissolved part of the volatile and is assumed to be incompressible and in thermomechanical equilibrium with a perfect gas phase, which is generated by the exsolution of the same volatile. An inverse problem approach is used -- the pressure along the duct is set as an input datum, and the other physical quantities are obtained as output. This fluid-dynamic model is intended as an approximate description of magma-volatile mixture flows of interest to geophysics and planetary sciences.
It is implemented as a symbolic code, where each line stands for an analytic expression, whether algebraic or differential, which is managed by the software kernel independently of the numerical value of each variable. The code is versatile and user-friendly and permits to check the consequences of different hypotheses even through its early steps. Only the last step of the code leads to an ODE problem, which is then solved by standard methods.
In the present work, the model is applied to the study of two sample cases, representing the ascent of magma-gas mixtures on the Earth and on a Jupiter's satellite, Io. Both cases lead to approximate but realistic descriptions of explosive eruptions, by taking pressure curves as inputs and outputting  conduit shapes together with mixture density, temperature and velocity along the ducts.

{\bf keywords:} {magmatic flow, mixture flow, compressible flow}
}

\section{Introduction}

A number of authors have modeled conduit flows of magma-volatile mixtures in order to represent the physical phenomena involved in volcanic eruptions.
These flows are studied in the cases of steady effusive and explosive volcanic activity 
(MC~GETCHIN \& ULLRICH 1973, WILSON et alii 1980, WILSON \& HEAD 1981, KIEFFER 1982, SLEZIN 2003), by using either of two conduit models to set boundary conditions in order that the system might be solved semi-analytically. More specifically, conduits are taken as being either parallel-sided or lithostatically pressure-balanced, the former model being characterized by walls parallel to one another and perfectly rigid 
(WILSON et alii 1980, WILSON \& HEAD 1981), the latter to a scenario where local changes in flow pressure relative to lithostatic pressure in the country rocks are promptly accommodated by wall failure 
(VALENTINE \& GROVES, 1996). A consequence of the parallel-sided scenario is the occurrence of the transonic transition of flow regime ($M=1$) at the surface, whereas the lithostatically pressure-balanced mode implies wallrock deformation and erosion, and assumes that the conduit is capable of adjusting itself such that stress across the walls is zero. 
In the pressure-balanced scenario, the flow regime transition is allowed to migrate downwards, from the surface to some depth in the upper conduit. As a result, upper-conduit velocities can be supersonic, but shocks and rarefaction waves may form at or after the transition zone 
(KIEFFER 1982, CHAPMAN 2000), which may cause massive changes in velocity, temperature and pressure in the fluid passing through them. Importantly, shock generation is very likely to be associated with non-gradual changes in conduit profile and/or abrupt variations in flow pressure gradients.

The present work is concerned with steady pressure-balanced scenarios, where the conduit shape evolution has come to an end and the conduit profile doesn't change with time. The presented model deals mainly with explosive volcanic activity, but allows for effusive activity as a subcase.
We model the ascent of magma and magma-gas mixtures from the base of the conduit up to the vent region. The magma rising through the crust is allowed to incorporate variable amounts of crustal volatile deposits. The volatile can be partly dissolved and partly exsolved (gas phase) in the magma, and for gas proportions larger than 10 wt\%, the mixture behaves as an ideal pseudogas 
(WALLIS 1969, KIEFFER 1982, LU \& KIEFFER 2009). Other model assumptions are taken from modelers of adiabatic flow of magma-gas mixtures through volcanic conduits 
(KIEFFER 1982, BURESTI \& CASAROSA 1989 and 1993, MASTIN 1995 and 2000), where no heat is transferred across the conduit walls during the eruption. The supersonic velocities resulting from the gain in kinetic energy occurring in pressure-balanced scenarios are counter-balanced by losses in elastic and thermodynamic energy.  

In the present model, the flow is one dimensional, steady and homogenous so, at any given depth, flow properties can be averaged across the entire cross-sectional area of the conduit. The gas phase, which is taken to be incondensable, behaves essentially as an ideal gas whereas the liquid phase is incompressible and in thermodynamic and mechanical equilibrium with the gas. Finally, at each depth the pressure of the mixture is taken to be uniform and equal to that of the gas phase alone.  
The model is completely analytic until it leads to an ordinary differential equation (or a set of ODEs in the most general case), which is then solved numerically by standard methods. This is obtained by suitable hypotheses, which simplify the theoretical treatment maintaining the capability of yielding predictions for the main physical quantities involved. 
These hypotheses, exposed in detail in the next sections, can be summarized as follows: the liquid phase (magma+dissolved volatile) is Newtonian; in this phase the dissolved mass fraction of volatile is in general much lower than the magma mass fraction; the exsolution process as a function of depth is slow; the conduit shape and the mixture temperature vary gradually; the mixture velocity exhibits large variations only in the upper, final part of the conduit. 

The corresponding author has written a Mathematica\copyright \, symbolic code - its advantage lying in its versatile tailorability for specific research needs and goals. Thanks to the symbolic approach, each line of the code is analytic like the corresponding equation of model, up to the last step, where a discrete computational component is still present, in that numerical methods for the solution of an ODE are used in computations. In this way, checks about the modification of an hypothesis, which may potentially give rise to a different model, can be easily done due to the fact that, as the code is running, many substitutions of an equation into a part of another, as well as derivations, integrations and so forth are being performed by the software. However, it is worth to remember that the appearance of non-physical or inconsistent calculations shall always be prevented by the author of the code.

Among the possible applications, two case studies of activity on the Earth and Jupiter's satellite, Io, are henceforth considered, though the analysis could be equally well-suited for any other planetary body in the Solar System for which evidence of present and/or past volcanic activity exists.

\section{Fundamental equations}

It is considered an unidimensional flow of a magma+volatile mixture in a vertical conduit, where the liquid fraction (magma+dissolved volatile) is  Newtonian and the exsolved volatile is a perfect gas, according to the hypotheses presented in the introduction above.
The conduit is extended over the vertical coordinate domain $z_0 \le z \le 0$, where $z=0$ is the conduit upper end at surface level (vent) and $z_0$ is the depth of the conduit lower end, the location of the latter being referred to as a {\em virtual reservoir}. This means that it may represent a real magma+volatile reservoir, or a zone where magma coming from deeper zones enters the conduit, with a chance of interacting with a given amount of crustal volatile deposits.
The conduit cross section $A$ is considered circular in what follows, $A(z)= \pi r^2(z)$ where $r(z)$ is the conduit radius, but this hypothesis is not so restrictive and can be easily generalized to different shapes. Similarly, the generalization from vertical to oblique conduits is not difficult, and it can be obtained by projecting the gravity acceleration on the direction of interest.
It is assumed that the variation of the conduit cross section is gradual, i.e. the radius varies slowly. A formal criterion for this property will be introduced in \S\ref{integr} and \ref{accur}. 

The equation set to consider in this problem is made of mass, momentum and energy conservation laws for a magma-gas mixture, and is written as follows: 
\bq
\lp( \rho\, v\, r^2 \rp)' = 0
\label{cont} 
\eq
\bq
p' + \rho g + f \frac {\rho v^2}{r} - 2\rho v^2\frac {r'}{r} - v^2 \rho' = 0
\label{mom}
\eq
\bq 
(c_p T)' + \lp( \frac{m}{\rho_m}\, p \rp)'+ v\, v' + g = 0
\label{ene}
\eq
where the prime denotes $z$-derivatives. Here $v(z),p(z),\rho(z),T(z)$ are velocity, pressure, density and temperature of the mixture,  $g$ is the gravitational acceleration and $f$ is a friction factor which can be related to the wall properties of the conduit (SCHLICHTING 1979). 
$f$ is typically constant for turbulent flows, but it could become $z$-dependent for low Reynolds numbers.
The energy equation involves also the constant pressure specific heat of the mixture $c_p(z)$, the magma mass fraction in the mixture $m(z)$ and the magma density $\rho_m$. In the most general case, all quantities depend on $z$ except $g$ and $\rho_m$.
The state equation of the mixture is written as 
\bq
p = \frac {\rho R T}{1 - m\, \rho/\rho_m} ,
\label{state}
\eq
where $R = n(z) R_g$ is a generalized form of the gas constant $R_g$, accounting for the gas mass fraction $n(z)=1-m(z)$. It represents the pressure equilibrium between magma and gas, and the denominator accounts for the volumetric fraction of magma. For large gas mass fractions, the state equation behaves as the pseudogas state formula $p = {\rho R T}+$ small corrections, which in turn tends to the ideal gas state law for $n \to 1$. 
Formally, this state equation and relations \rf{cont}...\rf{ene} form a system of 4 equations that characterize the problem under study.

Different approaches to the problem  are possible. Considering $g,f,\rho_m,m(z),c_p(z)$ as parameters of the model, to assign directly or to relate to the unknown functions by means of constitutive relationships, there are 5 unknown functions, namely $v(z),p(z),\rho(z),T(z)$ and $r(z)$. Starting from a formal system of 4 equations, two methods can be considered:

$\bullet$ Direct problem solution: the conduit shape is known by hypothesis together with the accessory conditions (function values at reference points, like the reservoir depth). The method leads to the flow pressure along the conduit and to the other functions $v(z),\rho(z),T(z),...$.

$\bullet$ Inverse problem solution: the flow pressure in the conduit $p(z)$ is known by hypothesis together with the accessory conditions. The method leads to the conduit shape as a radius curve $r(z)$ and to the other functions characterizing the flow.

The present work deals with an inverse problem, simplified by suitable assumptions, and implemented as a symbolic code, as disclosed in the introduction. 
A remarkable part of the analytic work is performed when the code is running until the desired result is achieved, i.e. until a final equation or set of equations is obtained, which is then solved numerically in the same code by means of high-level instructions.

\section{Model operation}
\label{moper}

\subsection{Starting data and lower conduit laws}
\label{sdata}

All formulas in what follows correspond to lines of the corresponding symbolic code. At first, as an input, a pressure curve $p(z)$ must be reasonably defined in a physically meaningful way on the vertical coordinate domain $z_0\le z \le0$.
Then, since the physics is remarkably influenced by the gaseous phase (exsolved volatile) present in the mixture, it must be considered that the present model deals with two main kinds of flow: if at a given depth $z_f$ the gas volume fraction $V_f$ of exsolved volatile contained in the mixture overcomes a critical value which is generally believed to be close to 0.75 (point of fragmentation), then the mixture velocity will increase dramatically, thus entering a compressible flow regime similar to that of ideal gases, with a chance for supersonic flow conditions to develop in the upper conduit and a potential to generate an explosive eruption. Alternatively, an effusive event will occur.

In the available literature, more than one critical value of $V_f$ is reported, which could all well suit the present model; in contrast, the formulation proposed here is not suitable for implementing fragmentation criteria based on the extensional-strain rate of a mixture element $dv/dz$.
Most importantly, in each particular case under study, it is not even known in advance whether or not fragmentation will occur at a certain $z=z_f$, 
which may be considered as a boundary region separating conduit portions characterised by different flow regimes. 
For this reason, the vertical domain must be considered in general as the union of two distinct domains and the pressure formally defined as the union of two separate curves,
\bq
p(z) = 
\begin{cases}
p_a(z) & $for $z_0 \le z \le z_f$ (lower conduit)$ \cr
p_b(z) & $for $z_f < z \le 0$  (upper conduit) .$ \cr
\end{cases}
\label{pr}
\eq
Here $p(z)$ is considered continuous across the whole domain $z_0\le z \le0$, but a discontinuity in its first derivative at $z_f$ is allowed.
This property may also appear in other relationships of this model without spoiling the validity of results, since the final equations can be solved over the two domains with suitable matching conditions.
On the other hand, the definition \rf{pr} may also be written in the form $p(z)=\theta(z_f-z) p_a(z)+\theta(z-z_f) p_b(z)$ where $\theta(z)$ is the step function, and the model can be easily generalized introducing a smoothing taper function $s(z-z_f)$, such as an hyperbolic tangent based on the conduit radius scale, and working with a pressure definition of the kind $p(z)=s(z_f-z) p_a(z) +s(z-z_f) p_b(z)$.
In the simplest cases, fragmentation doesn't take place and the "lower conduit" coincides with the overall domain.
Examples of initial guesses for pressure in realistic cases will be given in \S\ref{res}.

In this model, the fragmentation depth and pressure are obtained to a certain degree of accuracy. The total volatile mass fraction
$n_t$ must be set as an input datum, and may involve a number of volatile species, as follows:
\bq
n_t = \sum_i n_{ti} .
\label{ntot}
\eq
Each volatile species can be dissolved/exsolved in/from the magma in different amounts. For each species, models of dissolved fractions are found in the literature, and they all lead to the total fraction of dissolved volatile, as follows:  
\bq
n_d(z) = \sum_i n_{di}(z) = \sum_i F_i(z,...)
\label{ndtot}
\eq
where the symbols $F_i$ represent different dissolution models, which typically depend on $p(z)$ but may also affect one another (examples will be given in \S\ref{res}). Once the total dissolved fraction is known, the exsolved volatile in the magma, i.e. the gas mass fraction of the mixture, can be written as 
\bq
n(z) = n_t-n_d(z) ,
\label{netot}
\eq
and then the liquid (magma+dissolved volatile) fraction is $ m(z)=1-n(z) $.
The density $\rho_l$ of this liquid should account for the dissolved volatile mass fraction in the liquid, $n_v$, and the dissolved volatile density, 
$\rho_v$, but the relevant formula can be simplified as follows, under the general assumptions $n_v\ll 1$ and $\rho_v \ll \rho_m$:
\bq
\rho_l = \dfrac {\rho_{m} \rho_v}{(1-n_v) \rho_v + n_v \rho_m} \simeq \rho_m .
\label{ldens} 	
\eq

In this model the variations of $n_d(z),n(z)$ and $m(z)$ are taken to be very small and gradual, in order to neglect the relevant derivatives, a condition satisfied in many real cases. In a perturbative formulation based on a small parameter $\en$, this property would be expressed for example as $n=n(\en z)=n(Z)$ so that $dn/dz=\en\, dn/dZ$ and the calculations would be performed at order $O(1)$ by neglecting terms of order $O(\en)$.

Also, the total specific heat of any volatile species can be obtained by combining the heats of the different species 
(again, examples will be given in \S\ref{res}):
\bq
c_{pg}(z) = \frac {1}{n(z)} \sum_i n_i(z) c_{pgi} ,
\label{cpg}
\eq
where $n_i(z)=n_{ti}-n_{di}(z)$ is the exsolved fraction for the $i$-th species, and the specific heats $c_{pgi}$ are taken as mean values over a short range of temperatures around the lower end value $T_0$. This approximation relies on the hypothesis that, over a very wide range of eruption conditions studied in the literature, the temperature differences between the initial and final sections of the ducts are generally found to be limited to a few tens of degrees; in other words, the present model refers to flows where $\Delta T/T_0 \ll 1$.
In an analogous way, the specific gas constant for any gaseous volatile can be obtained as:
\bq
R_{g}(z) = \frac {1}{n(z)} \sum_i n_i(z) R_{gi} ,
\label{Rg}
\eq
where the $R_{gi}$ are the gas constants for the species involved.
Now, using the relation $c_{vg}=c_{pg}-R_g$ and accounting for the specific heat of the magma $c_m$ (input datum), the specific heats of the mixture at constant pressure and volume can be written as
\bq 
c_p(z) &=& n(z) c_{pg}(z) + m(z) c_m \label{cpm}\\
c_v(z) &=& n(z) c_{vg}(z) + m(z) c_m .\label{cvm}
\eq
As a consequence of the previous hypotheses, also $c_p,c_v$ and $R$ are slowly varying parameters.

To determine the fragmentation, it is necessary to know the gas density $\rho_g$, which is given here by the ideal law written for the lower conduit
\bq
\rho_{ga} = \frac{p_a}{R_g T_a} ;
\eq
this formula involves the temperature $T_a$ in the lower conduit, which is expressed in this model by integrating over $z$ the energy conservation 
\rf{ene}, starting from the lower end values $T_0,p_0,v_0$ at $z=z_0$ (input data), and simplifying the result under the hypothesis that in the lower conduit the velocity variations $(v_0^2-v^2)$ are small in comparison with the other terms in the original equation: 
\bq 
T_a(z) = T_0 + \frac{1}{c_p} \lp[ \frac{m}{\rho_m} (p_0 -p_a)+ \fh (v_0^2-v^2) + g\, (z_0-z) \rp]
\simeq T_0 + \frac{1}{c_p} \lp[ \frac{m}{\rho_m} (p_0 -p_a)+ g\, (z_0-z) \rp] .
\label{tapp}
\eq
Of course, the physical meaning of this approximation is that in the lower conduit the radius changes very slowly and the gas volume is small, which is generally true for a wide class of flows of this kind; the effects of this hypothesis can be checked as explained in \S\ref{accur}.
Then, the gas volume fraction can be written as
\bq
V_f = \frac{ n \rho_m}{n \rho_m + m \rho_{ga}} ,
\label{vfrac}
\eq
where the contribution of the dissolved volatile was neglected after equation \rf{ldens}. 
Finally, assuming that fragmentation takes place for a known critical value $V_{fc}$ (typically 0.75), and solving the equation $V_f(z_f)= V_{fc}$,
the fragmentation depth $z_f$ is obtained. The relevant pressure is $p_f=p_a(z_f)=p_b(z_f)$, so that the two domains (upper and lower conduits) are completely determined. Examples of initial guesses for pressure, as well as a check of the effects of the approximations introduced in this section will be given in \S\ref{res}. 

\subsection{Temperature and density on the complete domain}
\label{tdens}

Up to now, the temperature and density of the mixture have only been defined in the lower conduit. The treatment of the upper conduit in the most general case could turn out to be much more complicated, but in the present work it is simplified by focusing on the main physics involved in this region. An approximated and useful model for the temperature can be defined by considering that, after the fragmentation point $z_f$, the dominant phenomenon is  mixture expansion, and more specifically the expansion of the gas fraction as the magma is disrupted into small particles. 
Under the hypothesis of thermodynamic equilibrium, an isentropic expansion law for the mixture can be obtained as in 
BURESTI \& CASAROSA 1989, combining the isentropic condition $ds = c_p dT/T -R\, dp/p =0$ with the state law of the mixture \rf{state}, and considering the $z$-derivatives of $c_p$ and $R$ negligible as in the previous section. The resulting law can be expressed as 
\bq
p \lp(\frac{1 - m\, \rho/\rho_m} {\rho}\rp)^\gamma = const.,
\label{pexp}
\eq
or  
\bq
p^{\frac {1-\gamma}{\gamma}} T = const. ,
\label{texp}
\eq
where $\gamma$ is the ratio of the specific heats of mixture at constant pressure and volume \rf{cpm} and \rf{cvm}, 
\bq
\gamma(z) = c_p(z)/c_v(z) .
\label{gamma}
\eq
The considerations above suggest expressing the temperature in the following way:
\bq
T = \begin{cases}
T_a(z) = T_0 + \dfrac{1}{c_p} \lp[ \dfrac{m}{\rho_m} (p_0 -p_a)+ g\, (z_0-z) \rp] & $for $z_0 \le z \le z_f  \cr
T_b(z) = T_a(z_f) \lp[\dfrac {p_b}{p_f} \rp]^{\frac{\gamma(z) - 1}{\gamma(z)}}  & $for $z_f < z \le 0 , 
\label{temp}
\end{cases}
\eq
where the lower conduit law follows relation \rf{tapp} and the upper conduit law follows relation \rf{texp} representing an expansion starting from the value $T_a(z_f)$ and determined by the pressure variation $p_b(z)$ up to the final or vent pressure value $p_f$ (input data). 
Even if simplified, expression \rf{temp} for temperature doesn't prevent the appearance of phenomena like the adiabatic heating of the mixture, as will be shown in \S\ref{res}. 
Like $p(z)$, the definition \rf{temp} may have a discontinuity in its first derivative, which doesn't affect the problem solution. Also in this case, it can be removed by tapering.

The gas density $\rho_g$ in the upper conduit is still given by the ideal law, here $\rho_{gb} = {p_b}/(R_g T_b)$,
so that the global density can be written as:
\bq
\rho = \dfrac {\rho_{g} \rho_m}{m \rho_{g} + n \rho_m}  .	
\label{dens}
\eq
As the gas volume fraction \rf{vfrac}, this formula is simplified by neglecting the very small contribution of the dissolved volatile. 
Also in $\rho(z)$, a first derivative discontinuity at $z=z_f$ may be present, because it involves $p$ and $T$. This discontinuity is removable by tapering, and the same considerations as above hold for this function and all of the followings, when they involve $\rho$ and/or $p$, $T$.

\subsection{Velocity and Mach number, complete domain}

The flow speed can be related to the other functions by means of the continuity equation \rf{cont}. The velocity law turns out to be 
\bq
v(z) = \frac{\rho_0\, r_0^2\, v_0}{\rho(z)\, r^2(z)} ,
\label{vel}
\eq
where the input radius, velocity and density $r_0, v_0$ and $\rho_0$ are known as input data at the conduit lower end.
At this stage $v(z)$ is formally unknown, because $r(z)$ is unknown. This is not a problem for the symbolic code, which retains the
analytic expression of $v$ in the computer's memory.

The sound speed formula for the mixture, thanks to the hypothesis of thermodynamic equilibrium, can be obtained by relation \rf{pexp} using the definition
$c^2 =(\de p/\de\rho)_s$ and the state equation \rf{state}: 
\bq
c^2(z) = \frac {(\gamma R T)^{1/2}}{1 - m \rho/\rho_m} ,
\label{cs}
\eq
where $\gamma(z)$ is given by \rf{gamma} and $R = n(z) R_g$. As in the state equation, the denominator accounts for the volumetric fraction of magma. For large gas mass fractions, formula \rf{cs} expresses the sound speed in a pseudogas and finally, for $n \to 1$, in an ideal gas. 

The Mach number can be defined as usual, i.e. $M(z) = v(z)/c(z)$. As the velocity, $M(z)$ is at this stage known only in symbolic form. It is worth noting that, if supersonic flow develops, the sonic coordinate along the conduit can no longer be determined by the ideal condition $dr/dz=0$, as shown in the next subsection.

\subsection{Final integration}
\label{integr}

Up to now, continuity and energy conservation as well as relations having the physical meaning of a state law have been used in the model build-up. On the other hand, the momentum equation \rf{mom} relates all the quantities defined up to now and it can be used for the problem solution. 
If the variations of $\rho$ and $p$ are related to the speed of sound through relation $\rho' = (\de\rho/\de p)_s \,p' = p'/c^2$, the momentum equation can be rewritten as 
\bq
r'(z)- \dfrac{r(z)}{2 \rho(z) v^2(z)}\lp\{ p'(z) [1 - M^2(z)] + \rho(z) g + f \frac {\rho(z) v^2(z)}{r(z)} \rp\} = 0 .
\label{eqm}
\eq
Since the present method is completely analytic up to the integration of equation \rf{eqm}, by substitution of the previous formulas in this equation, a huge expression is obtained, which is far too cumbersome for reading. This expression is stored in the computer's memory by the symbolic code, and after substitution of the scalar input parameters, it can be solved for $r(z)$ by standard methods on the two domains $z_0 \le z \le z_f$ and 
$z_f < z \le 0$ with the boundary condition $r(z_0)=r_0$ and the continuity condition at $z=z_f$ (the continuity condition is no longer necessary if the pressure definition \rf{pr} has a continuous derivative over the whole domain).
In the symbolic code all steps (lines) but the last one are analytic. The last one instead provides the numerical solution, which is obtained through a single, high level, final instruction. The relevant numerical method can be automatically selected by the Mathematica\copyright\, software among the standard ones (Runge-Kutta, Adams...) in order to reach the highest level of accuracy in compliance with the internal representation of numbers in the computer used, and the integration step is selected in an adaptive way with the same goal. Alternatively, an experienced user can drive the solver in a more detailed way. 

Equation \rf{eqm} shows also that when $M=1$ the conduit radius satisfies the condition 

\centerline{$
r'= \frac 1 2 \lp[ \dfrac{r\, g}{ v^2} + f \rp]
$}

\nt
where the right hand side is always positive, so that the sonic condition can be reached only at positions where $r'>0$. This equation collapses into the ideal one if $g$ and $f$ vanish. 

The radius obtained by integrating equation \rf{eqm} must vary gradually in order to satisfy the basic hypotheses, and this can be expressed by the condition that the radius variation over a length scale in the order of the same radius must be small: 
\bq
|r(z+\Delta z)-r(z)|\ll r(z) \;\;\hbox{ for }\; \Delta z \sim r(z) \;\;\hbox{ and }\; z_0 \le z \le -r(0).
\label{slowr}
\eq

\subsection{Accuracy checks}
\label{accur}

The effect of the approximations introduced in the model can be checked {\em a posteriori} in different ways. A possible method relies on the use of energy conservation in integral form, 
\bq 
e_{t} =
c_p T +  \frac{m}{\rho_m}\, p + \frac 1 2 v^2 + g z
\label{econs}
\eq
which relates the total energy per unit mass $e_{t}$ to pressure, velocity and temperature curves. In particular, at the lower end depth, equation 
\rf{econs} reads

\centerline{$
e_0 =
c_{p0} T_0 +  \frac{m_0}{\rho_m}\, p_0 + \frac 1 2 v_0^2 + g z_0 .
$}

\nt
After the solution of the problem, i.e. the integration of equation \rf{eqm}, the relevant energy $e_t$ can be calculated by substituting in equation 
\rf{econs} the quantities $c_p,T,m,p$ and $v$ determined by the present model, and the resulting function $e_t(z)$ can be compared with the reference value $e_0$. For a "perfect model" $e_{t}$ should remain constant along $z$, i.e. $e_t=e_0$. In the present case the ratio
\bq
\left| \frac {e_0-e_t(z)}{e_0} \right|
\label{enest}
\eq
becomes an index of the method accuracy, representing the relative error on the total energy. As a general criterion, results could be validated if this index remains less than a given threshold, well below the unity. In fact, this quantity appears to be particularly sensitive to non-physical temperature variations: if the present model and code are used within the range set by the original hypotheses, the value of \rf{enest} amounts to less than a few percentage points, otherwise it grows rapidly. Similar indices can be defined for other quantities of physical interest.

A different but important validity check can be obtained by putting  condition \rf{slowr} in a linearized differential form, which gives rise to the quantity 
\bq
|r'(z)| ,
\label{rest}
\eq
i.e. a non-dimensional expression for radius variation rapidity.
In order to have a gradually varying duct, $|r'(z)|$ should remain well below unity along the conduit, and a criterion of comparison with a given threshold can be easily introduced. 
Another significant quantity is $|r'(z)/r(z)|$, which has the dimensions of a wavenumber. It may represents the scale of conduit shape fluctuations along $z$ and it should remain well below $1/r$.
Large values of $|r'(z)|$ distributed over long $z$ ranges may indicate a violation of unidimensionality, whereas large local variations of $|r'(z)/r(z)|$ may represent a violation of the isentropic assumption, such as that implied by the appearance of shocks in the compressible part of the flow.

\section{Sample results}
\label{res}

\subsection{Simplified formulas for flow pressure}
\label{pmodel}

For the sake of testing the method presented in section \S\ref{moper}, some examples of "pressure hypothesis" are given here.
In the simplest case, making the basic assumptions of incompressible flow in the subsonic range, the pressure could be taken as a linearly decreasing function; under conditions of constant mass fractions and heat coefficients, and slowly varying gas volume fraction, the same assumption would lead to slowly varying radii and velocities along the conduit, close to the extreme of pipe flow with friction where in the standard model $r$ and $v$ are constants. Thus, the simplest nontrivial hypothesis for the pressure is

\centerline{$
p(z) = p_0 - \dfrac {p_e - p_0}{z_0} (z - z_0)
$}

\nt
where the input data are the magma driving pressure $p_0$ of steady eruptions of magma and magma-gas mixtures (for which a plausible range of values can be chosen from available models) and the output pressure $p_e$ (known by models and/or direct observations). 

\begin{figure}
\begin{center}
\includegraphics[width=0.7\columnwidth]{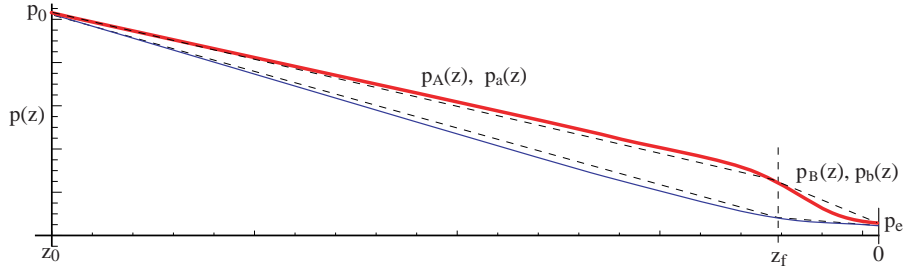}
\vspace{-5mm}
\caption{\small 
Sketch of real pressures compared with approximated pressures. Solid lines, thick and thin: examples of real pressure. Dashed lines: piecewise linearized pressures. The $z$ extent of the nonlinear zone in this figure is shown enlarged for clarity.
}
\label{ipres}
\end{center}
\end{figure}

However, the increase of the gas volume fraction has important effects -- if and when at a given depth $z_f$ the volume of exsolved volatile in the mixture overcomes the critical value and magma fragmentation occurs, even a pressure gradient which is linear at large depths could give rise to a compressible flow on a duct. 
In general, when a flow of this kind enters the compressible regime, there is also a deviation of the overall pressure curve from the linear behaviour; this deviation is often not negligible, even if it remains small in many interesting cases. Soon after magma fragmentation, the pressure typically decreases following a curve which is remarkably nonlinear only in the very upper conduit, this trend being qualitatively shown in figure \ref{ipres},  where the two solid curves (thick and thin) show different kinds of nonlinear behaviours. The $z$ extent of the nonlinear zone in this figure must be shown in an enlarged view for clarity, since often is very small with respect to the overall length of the conduit.
For a wide class of eruption flows, in the lower conduit, which represents the most vertically extensive portion of the duct, and according to 
{\small WILSON \& HEAD (1981)}, 
a real pressure curve in the lower conduit is expected to be similar to  the lithostatic pressure in the surrounding rocks because otherwise the conduit would be destroyed or heavily modified. The lithostatic pressure is typically very close to the linear behaviour or weakly nonlinear.
The property of $z_f$ to be very close to the end of the conduit, i.e. $|z_f| \ll |z_0|$, implies that the local deviation of $p(z)$ from a simple linear behaviour is not large, and in general the real flow pressure, a piecewise linearized approximation and the real lithostatic pressure are all close together, with the maximum differences becoming apparent only over a short (upper) portion of the conduit.

\begin{figure}
\begin{center}
\includegraphics[width=0.7\columnwidth]{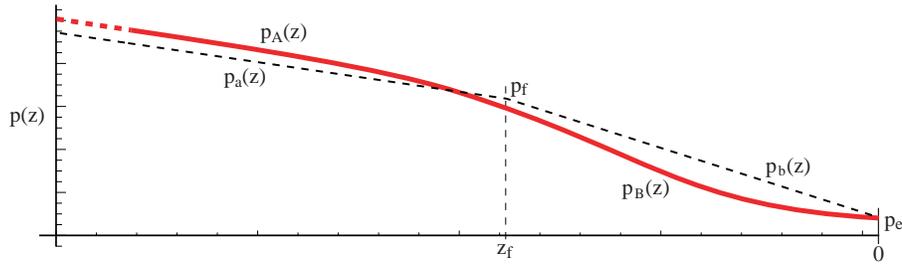}
\vspace{-5mm}
\caption{\small 
Particular of the approximated pressure curve, close-up in the fragmentation zone. Thick line: example of real pressure as in fig. \ref{ipres}. Dashed line: piecewise linearized pressure.
}
\label{zpres}
\end{center}
\end{figure}

For these reasons, here it is considered a piecewise linearization of $p(z)$ which introduces only a small error on the overall behaviour. 
The figure \ref{ipres} shows that in the following examples the real pressure $p(z)$, which can be considered as made of two parts $p_A(z)$ for $z\le z_f$ and $p_B(z)$ for $z > z_f$ (solid lines), can be approximated by piecewise linearized pressures $p_a(z)$ and  $p_b(z)$ in the same subdomains (dashed lines). Here the input data are still $p_0, p_e$ whereas the fragmentation depth and pressure $z_f$ and $p_f$ can be calculated as in \S\ref{sdata} with small errors. The details of real and approximated pressure curves for the thick line of figure \ref{ipres} in the fragmentation zone are sketched in figure \ref{zpres}. 


Formally, the approximated pressure curve proposed for the following examples is a piecewise linear function on the vertical coordinate domain 
$z_0\le z \le 0$:
\bq
p = 
\begin{cases}
p_a(z) = p_0 + \dfrac {\Delta p}{L} (z - z_0) 	 & $for $z_0 \le z \le z_f$ (lower conduit)$ \cr
p_b(z) = p_f - \dfrac {p_e - p_f}{z_f} (z - z_f) & $for $z_f < z \le 0$  (upper conduit)$ \cr
\end{cases}
\label{pinit}
\eq
The 1st derivative discontinuity can be avoided by tapering, as described in \S\ref{sdata}.
In equation \rf{pinit}, $p_0$ is the driving pressure at the lower end (depth $z_0$), which can be taken to be similar to the value of the lithostatic pressure for a fixed depth, with possible corrections accounting for varying crustal and tectonic settings across the planets/satellites under scrutiny;
$p_e$ is the output pressure at $z=0$, whilst $\Delta p$ is an estimation of the pressure drop along a conduit having a length in the order of $L=|z_0|$. This estimation can be obtained from the known law for pipes with friction, namely
\bq
\Delta p = - f \rho\, U^2 \frac {L}{2 D}
\label{deltap}
\eq
where $f$ is the friction factor, $\rho$ a scale density, which can be taken at this step equal to the magma density $\rho_m$, $U$ a scale velocity, in the order of the initial velocity parameter $v_0$, $D$ a scale diameter which can be taken equal to $2 r_0$ (twice the initial radius of the conduit). 

A code starting from equation \rf{pinit} can be implemented by defining as a first step $p_a(z)$ and following  the procedure of \S\ref{sdata} to determine the fragmentation depth $z_f$. Then $p_b(z)$ can be defined, and all the steps of \S\ref{tdens} to \ref{integr} can be followed.
If $z_f$ turns out to be 0 or greater, fragmentation doesn't take place, the lower conduit coincides with the whole duct, and equation \rf{pinit} collapses into the linear expression for $p_a(z)$. This is a model that suits an effusive eruption, in this case the input datum 
$p_e$ (exit or vent pressure) is not used, and the output pressure is automatically determined as the value of $p_a(0)$. This doesn't prevent the appearance of variations of radius and velocity, owing to the variation of gas volume fraction, but in general they are remarkably smaller than the variations in a flow with fragmentation.

Finally, it is worth to remember that the present model for pressure is just one of the possible and viable choices, and the symbolic code permits the exploration of different hypotheses in an easy fashion.
\subsection{An example based on terrestrial data.}
\label{earth}

As a first example, it is considered a mixture of terrestrial magma with CO$_2$ and H$_2$O. To set the volatile solubility properties, the magma is 
assumed to be basaltic, since basalt is the most common rock type found on the terrestrial planets. 
The input data at the conduit lower end are reported in what follows:

$z_0 = -5000$m ; $T_0 = 1350$K ; $p_0 = 7 \cdot10^7$Pa ; $\rho_m = 2900$kg/m$^3$ ; $r_0$ = 2m ; $v_0$ = 25 m/s .

\nt
In this example the input velocity represents a fully developed flow more than a flow starting from a real reservoir.
The total mass fractions of the volatile species are $n_{t{\rm CO_2}}$ =0.01 ; $n_{t{\rm H_2O}}$ =0.02 .
The volatile properties in this kind of magma are modeled after GERLACH (1986) 
and MASTIN (1995): 
the exsolved volatile mass fraction
$n_t$ is given by equation \rf{netot}, which reads here

\centerline{$
n(z) =  n_{{\rm CO_2}}(z) + n_{{\rm H_2O}}(z) = [n_{t{\rm CO_2}} - n_{d{\rm CO_2}}(z)] + [n_{t{\rm H_2O}} - n_{d{\rm H_2O}}(z)].
$}

\nt
The dissolved fractions for the two species are calculated in different ways, and in particular $n_{d{\rm CO_2}}$ is a linear function of the input pressure

\centerline{$
n_{d{\rm CO_2}}(z) = a_1 + a_2\, p(z) ,
$}

\nt
whereas $n_{d{\rm H_2O}}$ depends on $n_{d{\rm CO_2}}$ and must be obtained considering an algebraic system of the kind
\[
\begin{cases}
n_{d{\rm H_2O}} = F(p,p_p) & \cr
n_{d{\rm H_2O}} = n_{t{\rm H_2O}} - G(n_{d{\rm CO_2}},p,p_p)& \cr
\end{cases}
\]
where the complicated functions $F$ and $G$ are reported in the cited works 
(GERLACH 1986, MASTIN 1995) and also depend on the partial pressure $p_p$ of H$_2$O in the mixture. The solution of the system gives $n_{d{\rm H_2O}}(z)$ and $p_p(z)$. Then, the heat coefficients are obtained from \rf{cpg} and \rf{Rg}.
The remaining input data are $g = -9.81$m/s$^2$; $p_e = 10^5$Pa; $f = 0.06$. 
The value of friction factor $f$ represents a high roughness; here it is important to recall that the present definition of $f$ is consistent with the  pressure drop expressed by equation \rf{deltap}, as in SCHLICHTING (1979), 
but other authors use definitions which differ by a factor of four from the present one. 

\begin{figure}
\includegraphics[width=0.42\columnwidth]{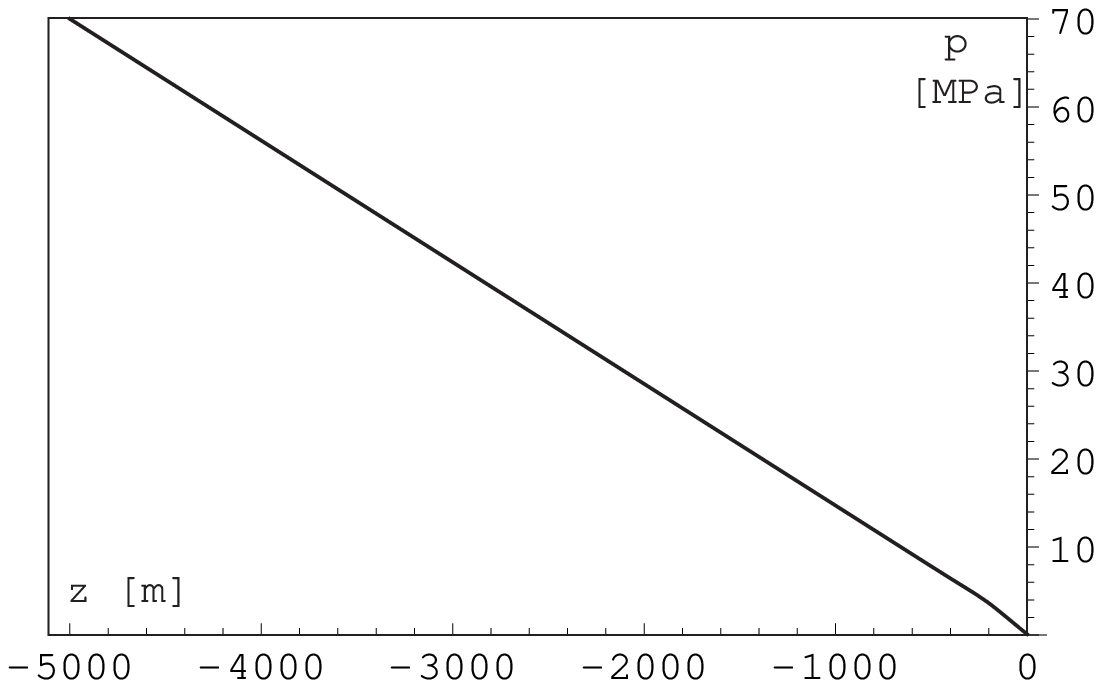}~a)
\hfil
\includegraphics[width=0.4\columnwidth]{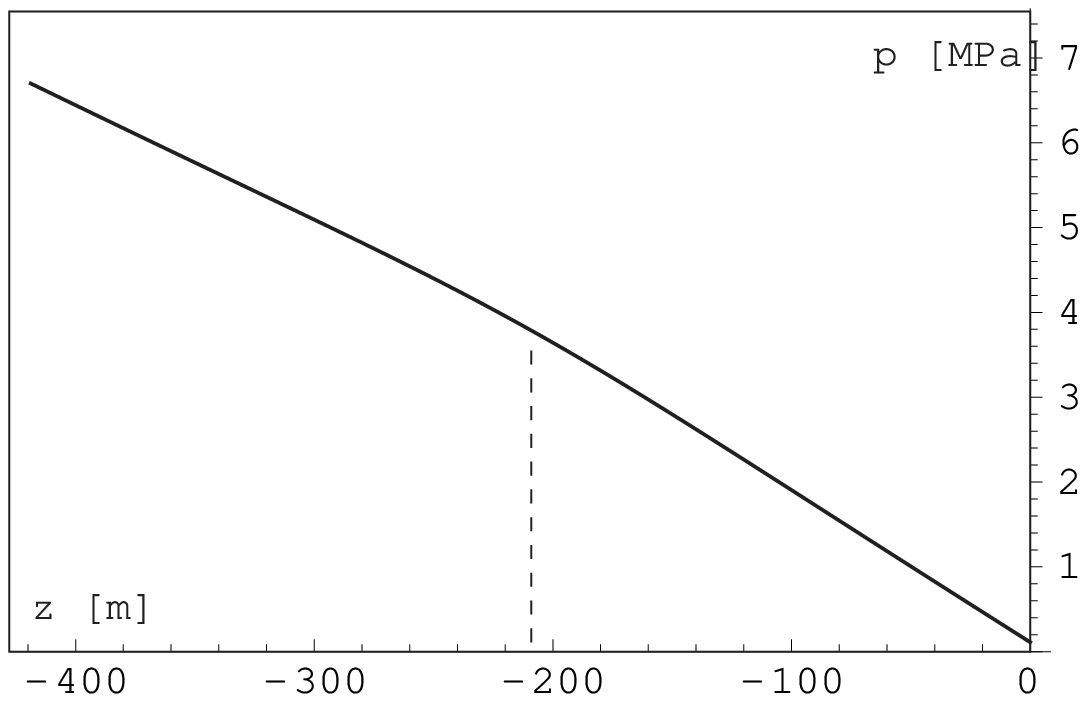}~b)
\caption{\small Terrestrial example, input pressure curve: overall \textbf{a} and close-up in the upper zone \textbf{b}. A weak knee is visible at the fragmentation depth, which is in this case -209m$\,\simeq 0.04 |z_0|$. The 1st derivative discontinuity is removed by tapering over a 50 diameters scale.}
\label{press1}
\end{figure}
\begin{figure}
\includegraphics[width=0.45\columnwidth]{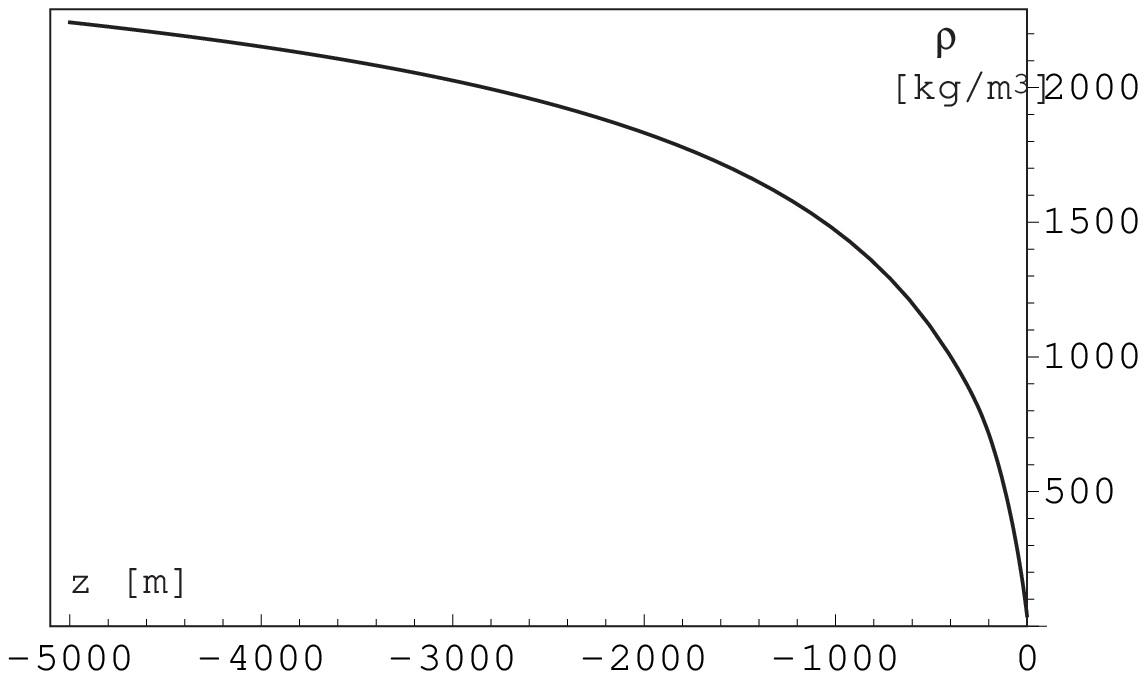}~a)
\hfil
\includegraphics[width=0.45\columnwidth]{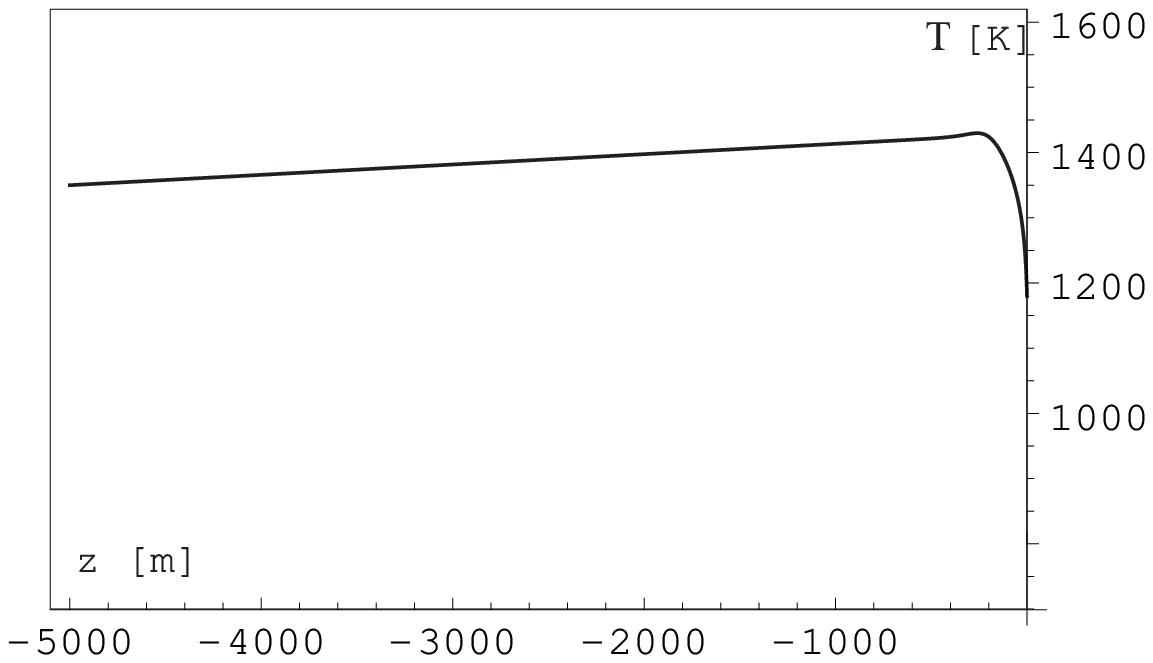}~b)
\caption{\small Terrestrial example, density  \textbf{a} and temperature \textbf{b}. The density exhibits a progressive decrease as the gas volume fraction increases. As said in the model description, the temperature is slightly raised by subsonic adiabatic heating, then decreases in the final expansion. The overall temperature variation ratio is $\Delta T/T_0\simeq 18\%$.}
\label{denstemp1}
\end{figure}

\begin{figure}
\includegraphics[width=0.4\columnwidth]{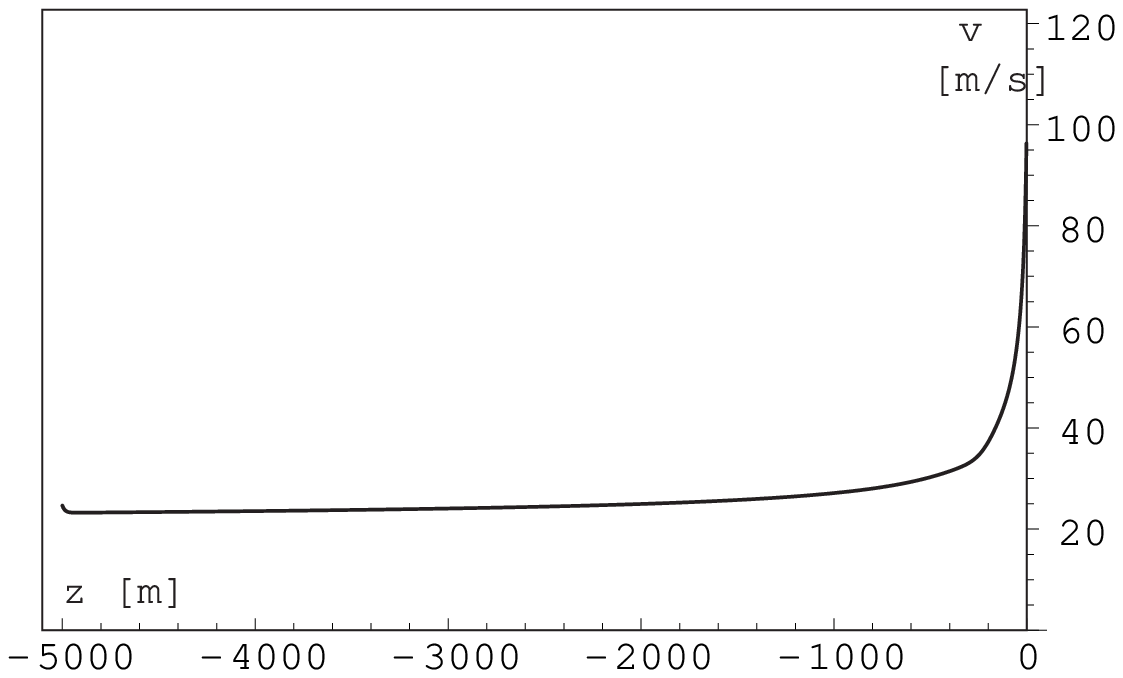}~a)
\hfil
\includegraphics[width=0.4\columnwidth]{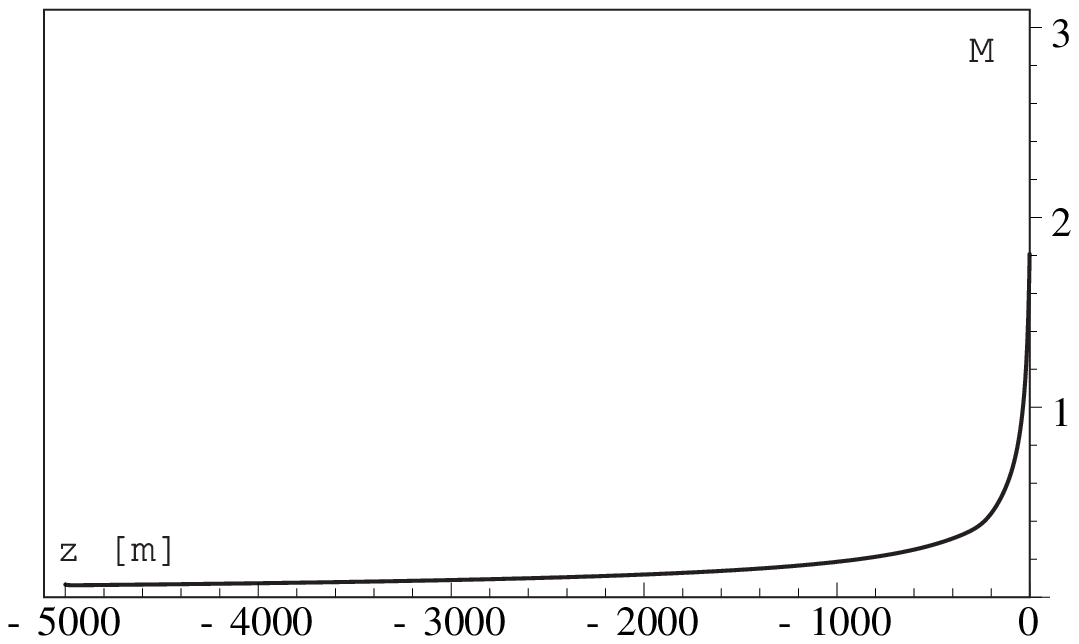}~b)
\caption{\small Terrestrial example, velocity \textbf{a} and Mach number \textbf{b}. The flow is sonic ($M=1$) at $z=-34.4$m. Output velocity is 
95.8m/s$\,\simeq 4 v_0$ and output Mach number is 1.81. This value of $M$ is related to the low value of the sound speed, which is in the mixture remarkably lower than in surrounding air.}
\label{velmach1}
\end{figure}

\begin{figure}
\begin{center}
\includegraphics[width=0.8\columnwidth]{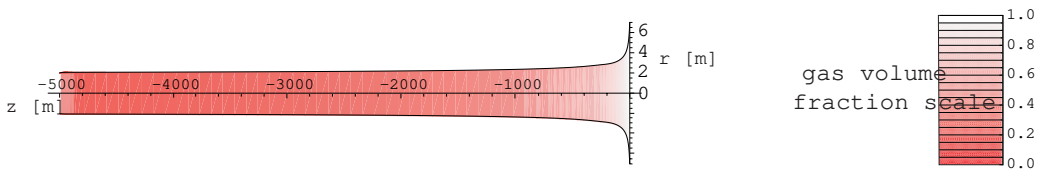}
\caption{\small Terrestrial example, conduit diameter (not to scale). White level is effectively proportional to gas volume fraction.}
\label{diam1}
\end{center}
\end{figure}

\begin{figure}
\begin{center}
\includegraphics[width=0.6\columnwidth]{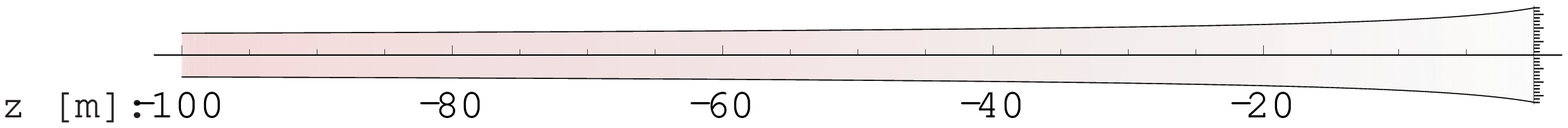}
\caption{\small Terrestrial example, conduit diameter, close-up in the output zone, scaled coordinates. White level proportional to gas volume fraction.
The output radius is 6.97m$\,\simeq 3.5 r_0$ and the final semidivergence angle is 18.0 degrees.}
\label{diam-z1}
\end{center}
\end{figure}

The input pressure curve modeled as in \S\ref{pmodel} is shown in figure \ref{press1} and the results are shown in figures \ref{denstemp1} to 
\ref{diam-z1}. A tapering was applied only to the input function $p(z)$. 
Output (vent) values of the physical quantities are reported in the captions for this particular case. The global error on the energy conservation after equation \rf{enest} is 2.4\%. The non-dimensional steepness after equation \rf{rest} is less than 0.024 up to 5 radii below the vent, and reaches the value 0.125 in the last radius below the vent. The scenario depicted in this example could represent an active lava fountain or the Plinian phase of an explosive eruption on the Earth.

It can be seen that the modification of few initial data, like the pressure profile, the total volatile mass fraction or the output pressure, leads to remarkable variations of the vent data such output velocity and divergence angle, which could represent different terrestrial eruptions.

\subsection{An example based on data for satellite Io.}

Another example, chosen to represent some properties of the eruptions on Jupiter's satellite, Io, is presented in what follows.  In the relevant literature, the magmatic materials are believed to be basaltic and/or ultramafics in composition, and the most important volatile species is assumed to be SO$_2$. This information is used here to set such necessary parameters, as the magma density and the volatile solubility properties. 
The input data at the conduit lower end are listed below:

$z_0 = -30000$m ; $T_0 = 1700$K ; $p_0 = 1.2 \cdot10^8$Pa ; $\rho_m = 3000$kg/m$^3$ ; $r_0$ = 0.42m ; $v_0$ = 6 m/s .

\begin{figure}
\includegraphics[width=0.42\columnwidth]{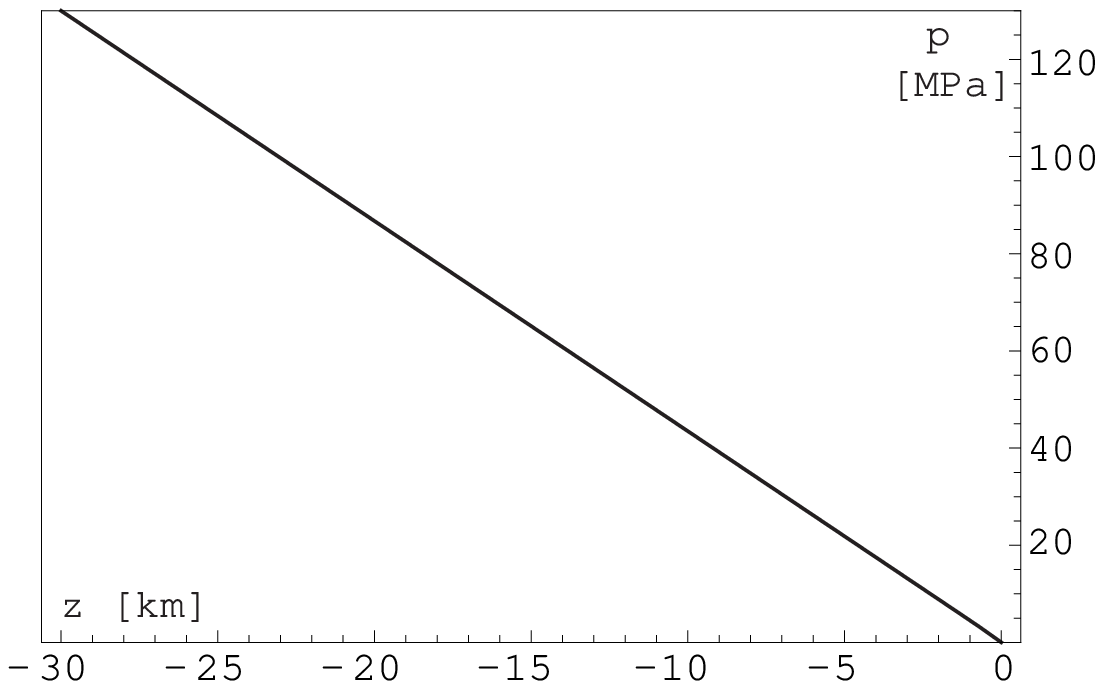}~a)
\hfil
\includegraphics[width=0.42\columnwidth]{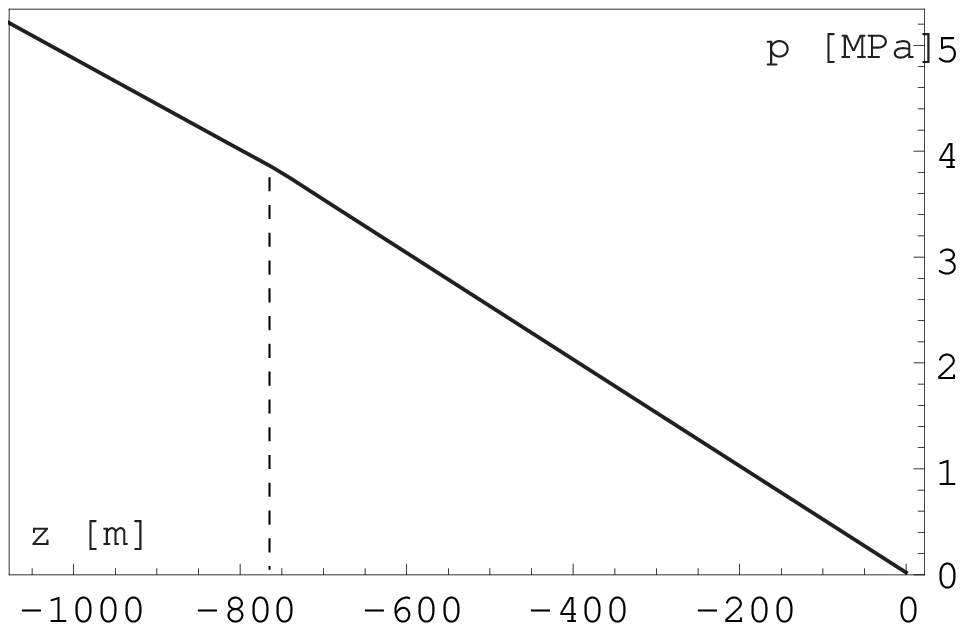}~b)
\caption{\small Example on Io, input pressure curve: overall \textbf{a} and close-up in the upper zone \textbf{b}. The fragmentation depth is in this case 
-774m$\,\simeq 0.026|z_0|$. The 1st derivative discontinuity is removed by tapering over a 75 diameters scale.}
\label{press2}
\end{figure}

\begin{figure}
\includegraphics[width=0.45\columnwidth]{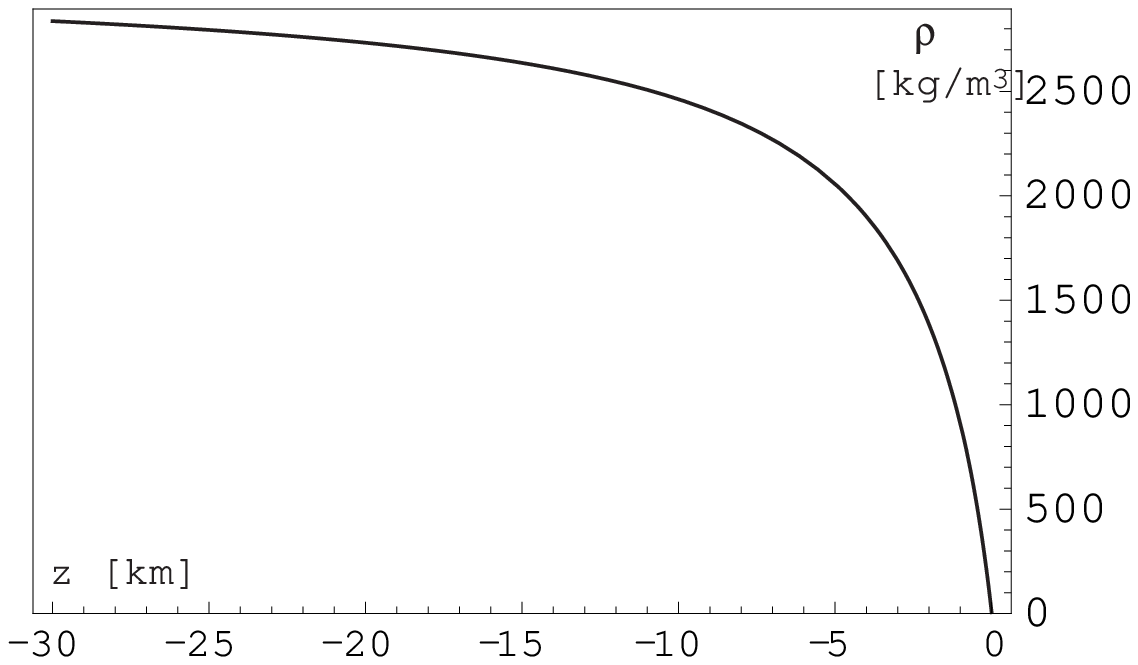}~a)
\hfil
\includegraphics[width=0.45\columnwidth]{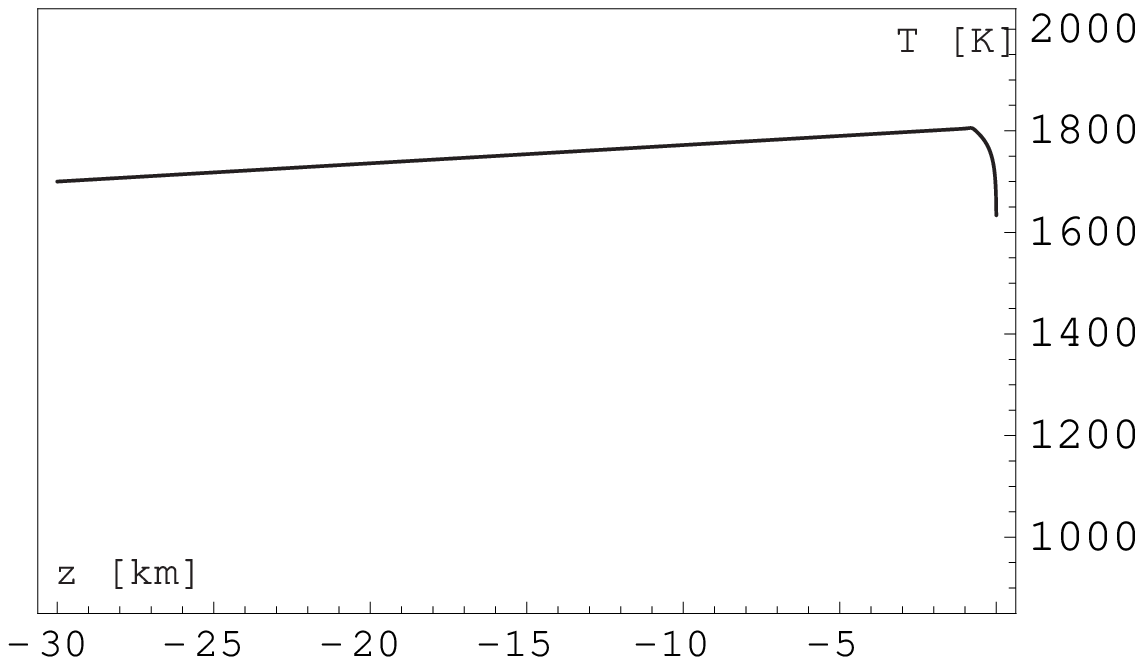}~b)
\caption{\small Example on Io, density  \textbf{a} and temperature \textbf{b}. The overall temperature variation ratio is $\Delta T/T_0\simeq 10\%$.}
\label{denstemp2}
\end{figure}

\begin{figure}
\includegraphics[width=0.4\columnwidth]{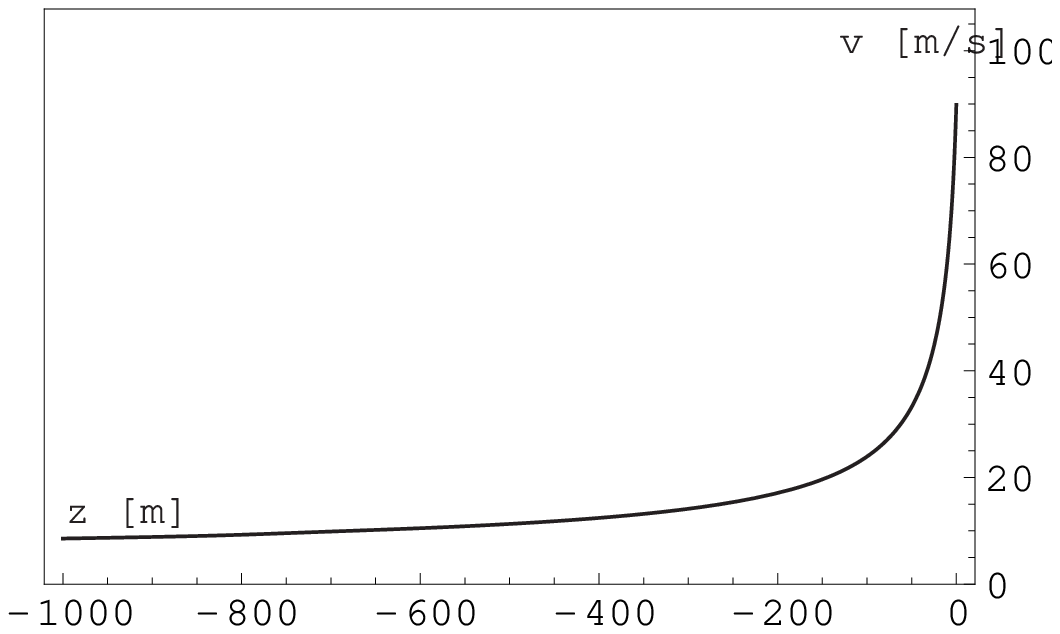}~a)
\hfil
\includegraphics[width=0.4\columnwidth]{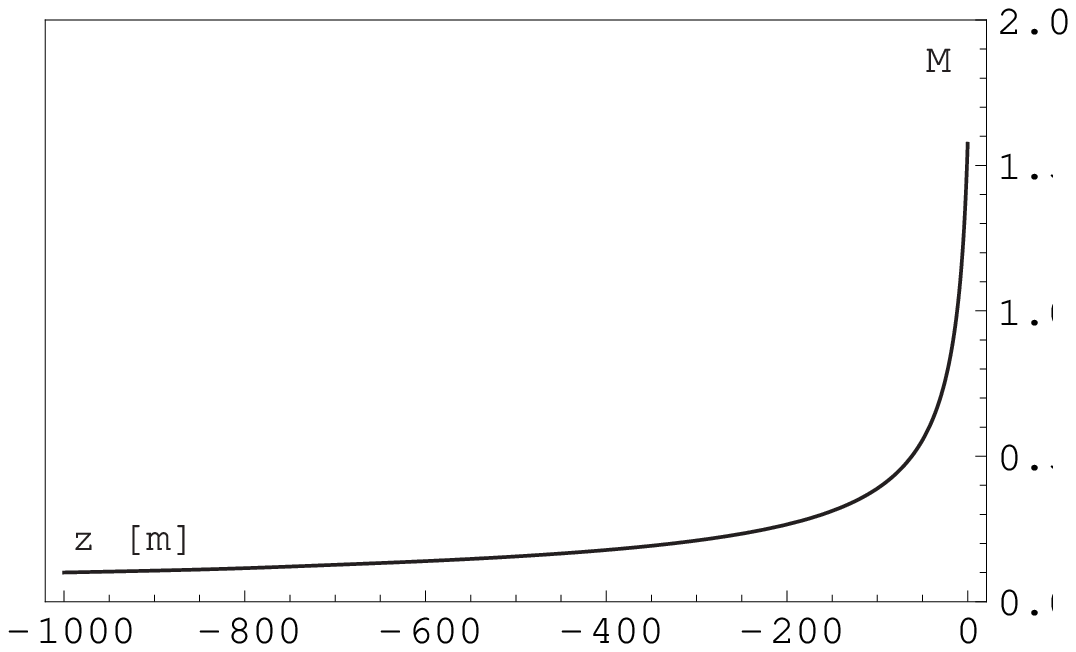}~b)
\caption{\small Example on Io, velocity \textbf{a} and Mach number \textbf{b} in the final part of the conduit. The flow is sonic at $z=-11.9$m. Output velocity is 89.9m/s$\,\simeq 14 v_0$ and output Mach number is 1.57. This value, which is related to the exit pressure chosen as an input datum, suggests the appearance of an underexpanded jet at the conduit vent since the surrounding ambient has a very low pressure. This is likely to represent the lowermost portion of a small-scale erupting plume on Io.}
\label{velmach2}
\end{figure}

\begin{figure}
\begin{center}
\includegraphics[width=0.6\columnwidth]{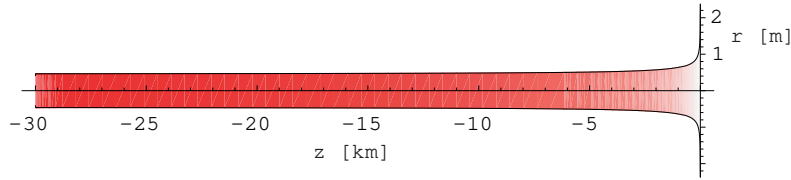}
\caption{\small Example on Io, conduit diameter (not to scale). White level is effectively proportional to gas volume fraction.}
\label{diam2}
\end{center}
\end{figure}

\begin{figure}
\begin{center}
\includegraphics[width=0.6\columnwidth]{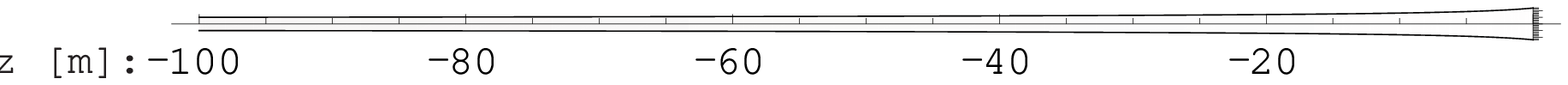}
\caption{\small Example on Io, conduit diameter, close-up in the output zone, scaled coordinates. White level proportional to gas volume fraction.
The output radius is 2.37m$\,\simeq 6 r_0$ and the final semidivergence angle is 10.5 degrees.}
\label{diam-z2}
\end{center}
\end{figure}

\nt
In this example the input velocity may represent the starting speed of the mixture after the initial development length at the reservoir-conduit connection. The total volatile mass fraction is $n_t=0.015$, which is likely to partly arise from the incorporation of gaseous/liquid SO$_2$ from shallow crustal aquifers, as shown in previous modelling of ionian eruptions 
(CATALDO et alii 2002). These crustal deposits are taken to be in proximity to $z_0$.
The volatile properties are modeled after MYSEN 1977, 
expressing the dissolved fraction $n_d(z)$ as a power series on $p$ of the kind

\centerline{$
n_d(z) = \sum_j a_j\, p^j ,
$}

\nt
approximated by a 3-d order truncation. The remaining input data are $g = -1.8$m/s$^2$; $p_e = 2 \cdot 10^4$Pa; $f = 0.055$.

The input pressure curve is shown in figure \ref{press2} and the results are shown in figures \ref{denstemp2} to \ref{diam-z2}. 
A tapering was applied only to the input function $p(z)$.
The global error on the energy conservation after equation \rf{enest} is in this case 1.9\%.
The non-dimensional steepness after equation \rf{rest} is less than 0.027 up to 5 radii below the vent, and becomes 0.099 at the last radius. 

\section{Conclusions}

The present paper deals with the study of the motion of magma-volatile mixtures in long vertical ducts with friction, originated from the desire to develop a model for the flow of magmatic fluid along volcanic conduits during steady phases of explosive eruptions, and allowing for effusive eruptions as a subcase. A simplified approach to this problem was taken, i.e. the mixture was assumed to be composed of an incompressible liquid phase and a perfect gas phase in conditions of thermomechanical equilibrium. The flow was treated as one-dimensional, homogeneous, steady and adiabatic. By means of further simplifying hypotheses, a "quasi-analytic" model capable of yielding predictions for the main physical quantities involved was obtained. In particular, the model is completely analytic through the whole series of steps except for the final ordinary differential equation (which becomes a set of ODEs in the most general case), that is solved numerically by standard methods.

The model is implemented as a symbolic code, where each line is similar to an equation of the model itself, or to what a human would write on a blackboard. Formal substitution of one equation into another and many other steps like derivations and integrations are performed by the software at running time, including the solution of the final ODE.
The code is easy to modify, offering the choice to change at any time the basic hypotheses of the model and/or introducing new ideas.
The degree of fulfilment of the assumptions upon which the model is based is checked by calculating suitable indicators {\em a posteriori}.

The model permits rapid analyses of the effects of the variation of the input parameters, a brief example was included for the sake of clarity, but without the aim of giving an exhaustive treatment, that could be the object of a further work.

Many further improvements in the model may be envisaged, like the study of non-circular cross sections, the introduction of different state equations, that could for example represent volatiles present in the mixture as supercritical fluids, and also the modification of the conservation laws in order to represent the interaction of a developed mixture flow with a crustal deposit of volatile that adds mass to the mixture in a given range of depths.
Again, these improvements could be considered as objects of future works.

\section*{References}

\footnotesize

BURESTI G. \& CASAROSA C. (1989) --
{\it One-dimensional adiabatic flow of equilibrium gas–-particle mixtures in long vertical ducts with friction.}
J. Fluid Mech. \textbf{203}, pp.\, 251--272. \\
%
BURESTI G. \& CASAROSA C. (1993) --
{\it The one-dimensional adiabatic flow of equilibrium gas–particle mixtures in variable-area ducts with friction.}
J. Fluid Mech. \textbf{256}, pp.\, 215-242. \\
%
CATALDO E., WILSON L., LANE S., GILBERT J. (2002) --
{\it A model for large-scale volcanic plumes on Io: Implications for eruption rates and interactions between magmas and near-surface volatiles.}
J. Geophys. Res. \textbf{107}, n. E11. \\
%
CHAPMAN C.J. (2000) -- 
High Speed Flow. Cambridge Univ. Press., Cambridge. \\
%
GERLACH T.M. (1986) -- 
{\it Exsolution of H2O, CO2, and S during eruptive episodes at Kilauea Volcano, Hawaii.} 
Journal of Geophysical Research, v. 91, no. B12, p. 12177-12185 .\\
%
KIEFFER S.W. (1982) --
{\it Fluid dynamics and thermodynamics of Ionian volcanism.} 
in: Satellites of Jupiter, edited by D.~Morrison (University of Arizona Press, Tucson) p. 647–-723.\\
%
LU X. \& KIEFFER S.W.  (2009) --
{\it Thermodynamics and mass transport in multicomponent, multiphase H2O systems of planetary interest.} 
Annual Reviews of Earth Planet. Sci., 37, 449--477.\\
%
MASTIN L.G. (1995) --
Open-File Report 95-756, 
U.S. Department of the Interior -- U.S. Geological Survey, Vancouver.\\
%
MCGETCHIN T.R. \& ULLRICH W.G. (1973)
{\it Zenoliths in maars and diatremes with inferences for the Moon, Mars and Venus. }
J. Geophys. Res. 78, 1833–-1853 .\\
%
MYSEN B.O. (1977) 
{\it Solubility of volatiles in silicate melts under the pressure and temperature conditions of partial melting in the upper mantle.} 
In: Proceed. Chapman Conf. on Partial Melting in the Upper Mantle, 
Portland, Oregon (Dept. Geol. Mineral. Industr, Portland), pp.\, 1--15. \\
%
SCHLICHTING H. (1979) -- 
Boundary-Layer theory.
MacGraw-Hill, New York.
%
SLEZIN Y.B. (2003) --
{\it The mechanism of volcanic eruptions (a steady state approach).} 
J. Volcanol. Geotherm. Res. 122, 7-–50.
%
VALENTINE G.A. \& GROVES K.R. (1996) --
{\it Entrainment of country rock during basaltic eruptions of the Lucero volcanic field,} 
New Mexico. J. Geol. 104, 71–-90. \\
%
WALLIS G.B. (1969) --
One-Dimensional Two-Phase Flow. MacGraw-Hill, New York.\\
%
WILSON L., SPARKS R.S.J., WALKER G.P.L. (1980) --
{\it Explosive volcanic eruptions — IV. The control of magma properties and conduit geometry on eruption column behaviour}
Geophys. J. Royal Astron. Soc. \textbf{63}, Issue 1, p.\,117 .\\
%
WILSON L. \& HEAD J.W. III (1981) --
{\it Ascent and Eruption of Basaltic Magma on the Earth and Moon.}
\it J. Geophys. Res. \textbf{86}, No.B4, p.\,2971 .\\

\end{document}